\documentclass[sigconf]{acmart}

\usepackage{graphicx}
\usepackage{textcomp}
\usepackage{xcolor}

\usepackage{bm}
\usepackage{algorithmic} 
\usepackage{algorithm}
\usepackage{tabularx}
\usepackage{subfigure}
\usepackage{booktabs} 
\usepackage{multirow}

\AtBeginDocument{%
  \providecommand\BibTeX{{%
    \normalfont B\kern-0.5em{\scshape i\kern-0.25em b}\kern-0.8em\TeX}}}

\copyrightyear{2023}
\acmYear{2023}
\setcopyright{acmlicensed}\acmConference[CIKM '23]{Proceedings of the 32nd
ACM International Conference on Information and Knowledge
Management}{October 21--25, 2023}{Birmingham, United Kingdom}
\acmBooktitle{Proceedings of the 32nd ACM International Conference on
Information and Knowledge Management (CIKM '23), October 21--25, 2023,
Birmingham, United Kingdom}
\acmPrice{15.00}
\acmDOI{10.1145/3583780.3615218}
\acmISBN{979-8-4007-0124-5/23/10}

\begin{document}
\fancyhead{}
\begin{sloppypar}  

\title{DPAN: Dynamic Preference-based and Attribute-aware Network for Relevant Recommendations}

\author{Wei Dai, Yingmin Su, Xiaofeng Pan}
\email{
    njuptdavid@163.com,guansuzai@126.com,pxfvintage@163.com
}
\affiliation{%
  \institution{Taobao \& Tmall Group}
  \country{China}
}

\author{Yufeng Wang}
\email{wfwang@njupt.edu.cn}
\affiliation{%
  \institution{Nanjing University of Posts and Telecommunications}
  \country{China}
}

\author{Zhenyu Zhu, Nan Xu, Chengjun Mao}
\email{{zzy234691, xiruo.xn, chengjun.mcj}@taobao.com}
\affiliation{%
  \institution{Taobao \& Tmall Group}
  \country{China}
}

\author{Bo Cao}
\email{zhizhao.cb@taobao.com}
\affiliation{%
  \institution{Taobao \& Tmall Group}
  \country{China}
}


\begin{abstract}
In e-commerce platforms, the relevant recommendation is a unique scenario providing related items for a trigger item that users are interested in. However, users' preferences for the similarity and diversity of recommendation results are dynamic and vary under different conditions. Moreover, individual item-level diversity is too coarse-grained since all recommended items are related to the trigger item. Thus, the two main challenges are to learn fine-grained representations of similarity and diversity and capture users' dynamic preferences for them under different conditions. To address these challenges, we propose a novel method called the Dynamic Preference-based and Attribute-aware Network (DPAN) for predicting Click-Through Rate (CTR) in relevant recommendations. Specifically, based on Attribute-aware Activation Values Generation (AAVG), Bi-dimensional Compression-based Re-expression (BCR) is designed to obtain similarity and diversity representations of user interests and item information. Then Shallow and Deep Union-based Fusion (SDUF) is proposed to capture users' dynamic preferences for the diverse degree of recommendation results according to various conditions. DPAN has demonstrated its effectiveness through extensive offline experiments and online A/B testing, resulting in a significant 7.62\% improvement in CTR. Currently, DPAN has been successfully deployed on our e-commerce platform serving the primary traffic for relevant recommendations. The code of DPAN has been made publicly available \footnote{https://github.com/DavidNeson/DPAN}.
\end{abstract}



\maketitle

\makeatletter
\renewcommand\@seccntformat[1]{\csname the#1\endcsname \hspace{0.5em}}
\makeatother

\section{Introduction}
\begin{figure}[htbp]
    \setlength{\abovecaptionskip}{0.1cm}
    \setlength{\belowcaptionskip}{-0.3cm}
    \centering
    \includegraphics[scale=0.7]{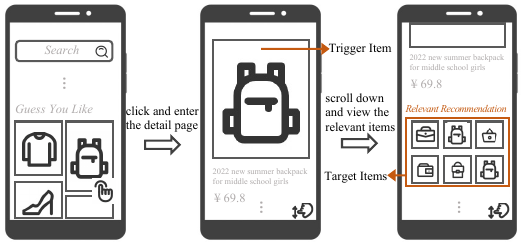}
    \caption{An illustration of the relevant recommendation of the item detail page on our online platform.}
    \label{fig:background}
\end{figure}

In e-commerce platforms, users’ instant interest can be explicitly induced with a trigger item, and follow-up related target items are recommended in relevant recommendations \cite{DIHN}.
As shown in Figure~\ref{fig:background}, users may click on a trigger item from different channels such as the Search Result Page (SRP) or Homepage Guess You Like (GUL), and then navigate to the detail page. As users scroll down to browse more information, relevant recommendations are displayed. The correlation between the target items and the trigger is ensured by matching algorithms. Our paper focuses on optimizing the Click-Through Rate (CTR) prediction of relevant recommendations.

With the progress of deep learning, CTR prediction has been studied from various aspects, including feature interaction \cite{DCN, DCNV2, DeepFM, xDeepFM}, sequential modeling \cite{DIN, DIEN, SIM, MOEF}, and dynamic neural networks \cite{Dynamic, M2M, APG, AdaSparse, FAN}. However, most of them neglect users’ interests in the trigger, leading to unsatisfactory user experiences if directly applied to relevant recommendations. Only a few studies have focused on joint modeling with the trigger. For example, R3S \cite{R3S} considers trigger articles' semantic relevance and information gain to provide users with more articles. DIHN \cite{DIHN} predicts the user’s real intent on the trigger and adaptively extracts instant interest from historical behaviors regarding the trigger and the target.

However, existing works overlook users' changing preferences for similarity and diversity in relevant recommendations under different conditions. Moreover, measuring diversity at the individual item level is too coarse as all target items are related to the trigger. For instance, when users click from SRP, they may prefer items with the same shop and category or similar images and titles to the trigger, indicating a preference for similarity. Conversely, when users click from GUL, they may prefer diversity and choose items with different attributes, such as different categories or brands compared to the trigger. Hence, modeling faces two challenges: learning attribute-aware representations of similarity and diversity, and capturing users’ dynamic preferences under various conditions.

We propose the Dynamic Preference-based and Attribute-aware Network (DPAN) to tackle the above challenges. Firstly,  DPAN leverages rich attribute information (e.g., category, brand, price, and title) to generate activation values for each attribute sequence via Attribute-aware Activation Values Generation (AAVG). Then, Bi-dimensional Compression-based Re-expression (BCR) is employed after AAVG to obtain fine-grained similarity and diversity representations of user interests and item information. As users' intentions (e.g., buying or browsing) may change under different conditions, leading to varying preferences for similarity and diversity of recommendations, we propose the Shallow and Deep Union-based Fusion (SDUF) to capture users' dynamic preferences adaptively. 

Our main contributions are summarized as follows:

$\bullet$ We propose a novel CTR model named DPAN to satisfy users’ dynamic preferences for similarity and diversity of relevant recommendations through learning attribute-aware representations.

$\bullet$ We design the Bi-dimensional Compression-based Re-expression to obtain the similarity and diversity representations of user interests and item information through mining information at the attribute level, and the Shallow and Deep Union-based Fusion to capture users’ dynamic preferences in varying conditions.

$\bullet$ We conduct extensive offline experiments and online A/B testing in the industrial environment. Our results demonstrate the superior effectiveness of DPAN compared to SOTA models.

\section{proposed method} \label{sec:method}

\subsection{Overall Architecture}
\begin{figure*}[htbp]
    \setlength{\abovecaptionskip}{0.1cm}
    \setlength{\belowcaptionskip}{-0.4cm}
    \centering
    \includegraphics[width=1\linewidth]{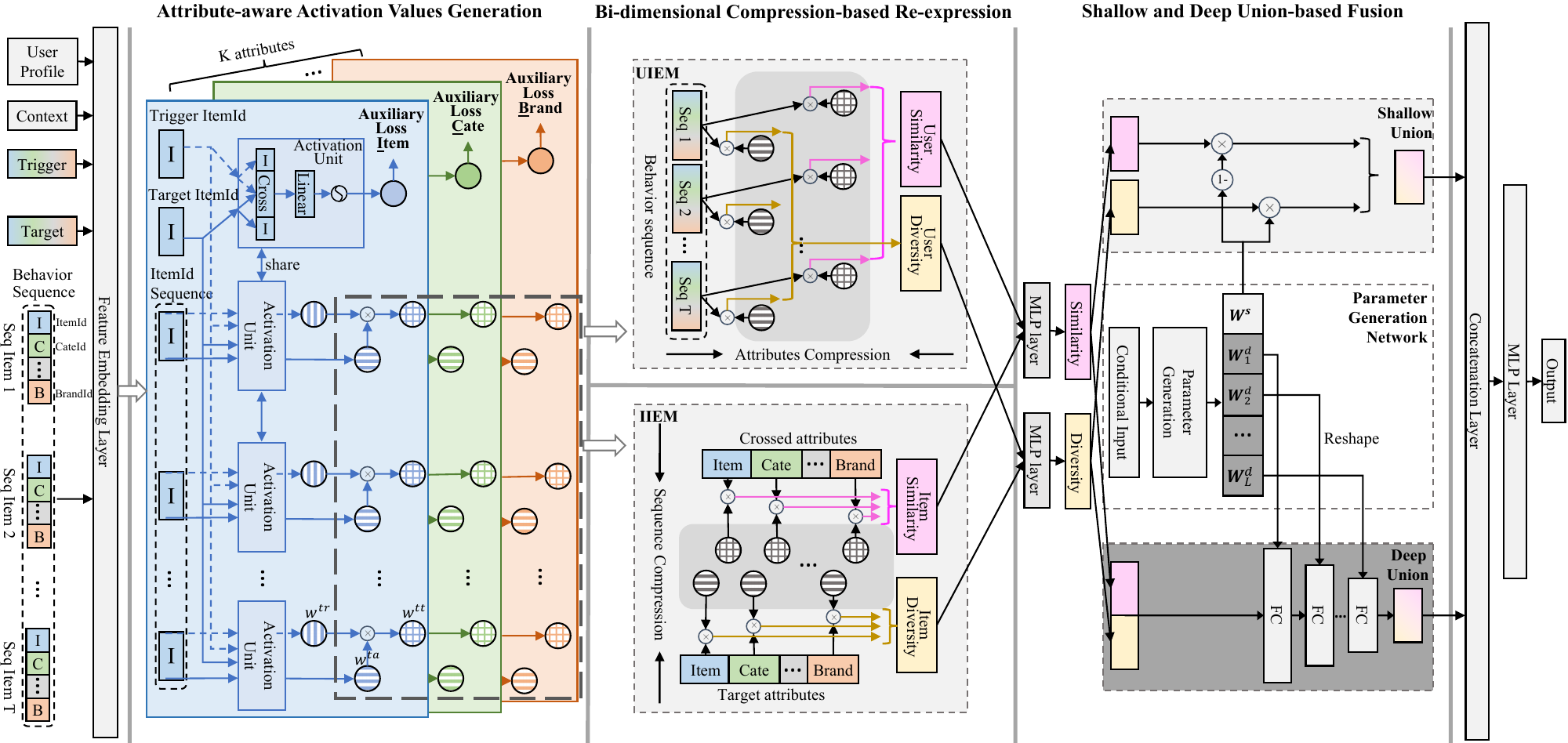}
    \caption{The overall architecture of DPAN. The core methods comprise Attribute-aware Activation Values Generation (AAVG), Bi-dimensional Compression-based Re-expression (BCR), and Shallow and Deep Union-Based Fusion (SDUF).}
    \label{fig:DPAN}
\end{figure*}

Figure~\ref{fig:DPAN} illustrates the overall architecture of DPAN. Firstly, all input features are transformed into embeddings by a feature embedding layer, including: 1) trigger item $\bm{e}^{tr}$=$[\bm{a}^{tr}_1,...,\bm{a}^{tr}_K]$, where $\bm{a}^{tr}_k$, $\forall k \in [1, K]$ denotes the embedding of the trigger's $k$-th attribute, and $K$ is the number of attributes; 2) target item $\bm{e}^{ta}$=$[\bm{a}^{ta}_1,...,\bm{a}^{ta}_K]$; 3) user behavior sequence $\bm{e}^{s}=\{\bm{e}_1^{s},...,\bm{e}_T^{s}\}$, where $\bm{e}_i^{s}=[\bm{a}_{i1}^{s},...,\bm{a}_{iK}^{s}]$, $\forall i \in [1, T]$ denotes the $i$-th behavior in sequence and $T$ is the truncated sequence length; 4) user profile $\bm{e}^{u}$; and 5) context information $\bm{e}^{c}$, such as the trigger's located channel, users' browsing time, and promotion events \cite{Metacvr, MOEF}. Subsequently, the behavior sequence is split into several attribute sequences. Each attribute sequence is activated by the trigger and the target respectively, generating corresponding activation values through Attribute-aware Activation Values Generation (AAVG). Then AAVG output is processed using Bi-dimensional Compression-based Re-expression (BCR) to learn similarity and diversity representations of user interests and item information. Finally, with the aggregated representations of similarity and diversity obtained through a Multi-Layer Perceptron (MLP) as input, dynamic preference results are obtained via Shallow and Deep Union-based Fusion (SDUF) under various conditions.

\subsection{\resizebox{3.08in}{!}{Attribute-aware Activation Values Generation}}
Firstly, $\bm{e}^{s}$ is divided into $K$ attribute sequences, denoted as $\bm{a}^s_k=\{\bm{a}_{1k}^{s},...,\bm{a}_{Tk}^{s}\}$ for the $k$-th attribute. Then, $\bm{a}_{ik}^{s}$, the embedding of the $i$-th element in $\bm{a}^s_k$, is activated by $\bm{a}^{tr}_k$ and $\bm{a}^{ta}_k$ respectively via DIN's \cite{DIN} activation unit to calculate corresponding activation values, namely $w^{tr}_{ik}$ and $w^{ta}_{ik}$. All activation units share the same structures and parameters for the same attribute, and the Hadamard product is adopted for attribute crossing. We also feed $\bm{a}^{tr}_k$ and $\bm{a}^{ta}_k$ into the activation unit to calculate $\hat{y}_k$, predicting whether the $k$-th attribute is clicked. Accordingly, we formulate the auxiliary loss as:
\begin{equation} \label{eq:aux_loss}
    \setlength\abovedisplayskip{2pt}
    \setlength\belowdisplayskip{2pt}
    Loss^{aux}_k = - \frac{1}{|\mathcal{D}|} \sum\nolimits_{(\bm{x},y) \in \mathcal{D}} (y\,log\,\hat{y_k} + (1-y)\,log(1-\hat{y_k}))
\end{equation}
where $\mathcal{D}$ is the training set with $|\mathcal{D}|$ samples, with $\bm{x}$ as the input and $y \in \{0,1\}$ as the label of whether the target is clicked. The auxiliary loss for each attribute is added to the final loss with a zoom factor $\alpha$.
Further, we calculate the dual activation score as $w^{tt}_{ik}=w^{tr}_{ik} \cdot w^{ta}_{ik}$, which will be high if $\bm{a}_{ik}^{s}$ is highly correlated with the corresponding attribute of the trigger and the target simultaneously. 

\subsection{\resizebox{3.08in}{!}{Bi-dimensional Compression-based Re-expression}}
After AAVG, we devise Bi-dimensional Compression-based Re-expression (BCR) to learn similarity and diversity representations of user interests and item information at the attribute level. To elaborate, we compress along the attribute dimension in User Interests Embedding Module (UIEM) to derive user interests representations and compress along the sequence dimension in Item Information Embedding Module (IIEM) to capture item information representations.

\textbf{UIEM}. 
The overall correlation between the target and each sequence item is obtained by averaging the target-activated scores of each attribute, i.e., compressing along the attribute dimension. UIEM focuses on the user behaviors with higher correlation scores and obtains the representation of user interests towards the target via weighted sum pooling. Behaviors with higher relevance to the target receive higher weights and dominate the representation of user interests. Thus the diversity representation of user interests, which varies over different target items, is calculated as follows,
\begin{equation} \label{eq:u_div}
    \setlength\abovedisplayskip{2pt}
    \setlength\belowdisplayskip{2pt}
    \bm{v}_U^{div}=\sum\nolimits _{i=1}^T(\frac1K \sum\nolimits _{k=1}^K w^{ta}_{ik})\bm{e}_i^{s}
\end{equation}

In order to obtain the similarity representation of user interests, UIEM captures items that are related to both the trigger and the target in the behavior sequence, so the target-activated scores are replaced by dual-activated scores, which are calculated as follows:
\begin{equation} \label{eq:u_sim}
    \setlength\abovedisplayskip{2pt}
    \setlength\belowdisplayskip{2pt}
    \bm{v}_ U^{sim}=\sum\nolimits _ {i=1}^T(\frac1K \sum\nolimits _{k=1}^K w^{tt}_{ik})\bm{e}_i^{s}
\end{equation}
It is noteworthy that only behaviors that are related to both the trigger and the target at the same time will receive higher weights, thus ensuring the similarity between user interests and the trigger.

\textbf{IIEM}. 
Users have varying levels of attention toward different attributes. To determine the user’s preference score for each attribute, the target-activated scores of all items in each attribute sequence are averaged, i.e., compressing along the sequence dimension. IIEM assigns higher weights to attributes with higher scores and re-expresses the target via weighted sum pooling,
\begin{equation} \label{eq:i_div}
    \setlength\abovedisplayskip{2pt}
    \setlength\belowdisplayskip{2pt}
    \bm{v}_I^{div}=\sum\nolimits _ {k=1}^K(\frac1T \sum\nolimits _{i=1}^T w^{ta}_{ik})\bm{a}^{ta}_k
\end{equation}
Note that with the trigger information excluded, Equation~\ref{eq:i_div} re-express the target according to the rich user behaviors, making $\bm{v}_I^{div}$ a diversity representation of item information. 

For similarity, we take into account users’ focus on both the trigger and the target attributes. We replace the target-activated scores with dual-activated scores and assign weights to the crossed attributes of the trigger and the target, as shown below:
\begin{equation} \label{eq:i_sim}
    \setlength\abovedisplayskip{2pt}
    \setlength\belowdisplayskip{2pt}
    \bm{v}_I^{sim}=\sum\nolimits _ {k=1}^K(\frac1T \sum\nolimits _{i=1}^T w^{tt}_{ik})cross(\bm{a}^{tr}_k, \bm{a}^{ta}_k)
\end{equation}
These crossed attributes are shared with the crossed attributes in the activation unit. The similarity representation of item information incorporates both the trigger and the target information, while also being influenced by users' preferences for different attributes.

\subsection{Shallow and Deep Union-based Fusion}
Following BCR, the similarity and diversity representations of user interests and item information are combined via different MLPs to generate the aggregated similarity and diversity representations with the same dimension, denoted as $\bm{v}^{sim}$ and $\bm{v}^{div}$. To capture users' dynamic preferences, we designed Shallow and Deep Union based-Fusion (SDUF), which includes a shallow union network, and a deep union network of $L$ Fully-Connected Layers. The aggregated representations are fed into these networks to generate fusion representations in both low-dimensional and high-dimensional spaces. The parameters of these two networks are dynamically generated through a parameter generation network \cite{APG, AdaSparse}.

To generate parameters for the union networks, we input conditional information $\bm{h}_{cond}$ (e.g., context and trigger information) into the parameter generation network. Note that target information cannot be used as conditional input since it represents posterior knowledge. Then a series of parameters is generated as follows:
\begin{equation}
    \setlength\abovedisplayskip{2pt}
    \setlength\belowdisplayskip{2pt}
    \bm{W}=\mathcal{PG}(\bm{h}_{cond})=MLP(\bm{h}_{cond})\Rightarrow [\bm{W}^s, \bm{W}_1^d, \bm{W}_2^d,...,\bm{W}_L^d]
\end{equation}
where $\bm{W}^s$ denotes parameters of the shallow union network, and $\bm{W}_1^d, \bm{W}_2^d,...,\bm{W}_L^d$ are parameters for the deep union network. 

The shallow union network combines the similarity and diversity representations using the element-wise product with $\bm{W}^s$ as follows:
\begin{equation}
    \setlength\abovedisplayskip{2pt}
    \setlength\belowdisplayskip{2pt}
    \bm{v}^{su}=\bm{W}^ s \otimes \bm{v}^{div}+(1-\bm{W}^s) \otimes \bm{v}^{sim}
\end{equation}

In the deep union network, $\bm{v}^{sim}$ and $\bm{v}^{div}$ are concatenated as input $\bm{h}_0=[\bm{v}^{sim}, \bm{v}^{div}] \in \mathbb{R}^{d_0}$. We reshape $\bm{W}_l^d$ to obtain $l$-th layer's parameters, i.e., $\bm{W}_l=reshape(\bm{W}_l^d) \in \mathbb{R}^{d_{l-1} \times d_{l}}$. The $l$-th layer's output is calculated as $\bm{h}_l = \sigma (\bm{h}_{l-1} \bm{W}_l)$, and we use the $L$-th layer's output as the result of the deep union network, i.e., $\bm{v}^{du}=\bm{h}_L$.

Finally, the results of SDUF, along with input features (i.e., $\bm{e}^{tr}$, $\bm{e}^{ta}$, $\bm{e}^u$ and $\bm{e}^c$), are combined and fed into the final scoring network through an MLP with Sigmoid activation for resulting click probability. The widely-used logloss is used for training, and auxiliary losses in Equation~\ref{eq:aux_loss} are also incorporated into the final loss.

\section{experiments}

\subsection {Experimental Setup}

\textbf{Dataset}. 
To the best of our knowledge, there are no public datasets available for relevant recommendations. Therefore, we obtained our proprietary industrial dataset by collecting logs from our platform's relevant recommendation scenario between 05/01 and 05/14, 2022. To mitigate the impact of invalid impressions caused by users quickly scrolling through the page, we treat clicked items as positive samples, while non-clicked items around the clicked items as negative samples in sample selection. The detailed statistics are shown in Table~\ref{tab:dataset}. Then we split the dataset into two non-overlapping parts for training (05/01-05/13) and testing (05/14). 

\textbf{Competitors}. 
\textbf{DIN} \cite{DIN} and \textbf{DIEN} \cite{DIEN} aim to extract users' (evolving) interests with respect to the target. \textbf{DHAN} \cite{DHAN} captures users' interests in different attributes through multi-dimensional hierarchical structures. To ensure fair comparisons, we incorporate the trigger information into the above methods via the activation calculation between the trigger and users' historical behaviors. In relevant recommendations, R3S \cite{R3S} effectively extracts information between the target and the trigger, but it neglects the modeling of sequential behaviors. Thus, we implement \textbf{R3S+DIN} for comparison. \textbf{DIHN} \cite{DIHN} fuses the trigger and the target embeddings based on the users’ intent on the trigger and then extracts users' interests from historical behaviors through hybrid interest extraction.

\begin{table}[t]
\centering
    \caption{Statistics of the collected dataset}
    \vspace{-2mm}
    \setlength{\tabcolsep}{2mm}{
                \begin{tabular}{l c c c c}
                \hline
                \#Users  &\#Triggers & \#Items   & \#Impressions & \#Clicks \\
                \hline
                2835644 & 3321513 & 8262925  & 59355035     & 13668100  \\
                \hline
            \end{tabular}
    }
    \label{tab:dataset}
\vspace{-2em}
\end{table}

\textbf{Experimental Settings}. 
We implemented all methods using TensorFlow 1.12, with a mini-batch size of 1024, and Adagrad optimizer with a learning rate of 0.01. 
For DPAN, we set the zoom factor $\alpha$ to 0.1, the truncated user sequence length $T$ to 50, and the embedding size for each attribute to 16. 
The attributes used include item, brand, category, price, and title. The dimension of aggregated representations is 128. The conditional input consists of the channel, browsing time, and trigger information.
In the deep union network, $L$ is set to 2 and the number of layers is [256, 128]. We take AUC as the evaluation metric and use RelaImpr \cite{RelaImpr} to measure relative improvements.

\vspace{1ex}
\noindent
\hspace{0.2em}
\begin{minipage}[htbp]{\columnwidth}
\begin{minipage}[htbp]{0.48\columnwidth}
\centering
\makeatletter\def\@captype{table}\makeatother\caption{Offline Comparison}
\vspace{-3mm}
\resizebox{!}{0.55in} {
    \begin{tabular}{l c c}
    \hline
    Model   & AUC     & RelaImpr \\ 
    \hline
    DIN     & 0.5992  & 0.0\%    \\
    DIEN    & 0.6009  & 1.71\%   \\
    DHAN    & 0.6025  & 3.33\%   \\
    R3S+DIN & 0.6037  & 4.54\%   \\
    DIHN    & 0.6052  & 6.05\%   \\
    \hline
    \textbf{DPAN} & \textbf{0.6098} & \textbf{10.69\%} \\
    \hline
    \end{tabular}
    }
\end{minipage}
\hspace{0.1cm}
\begin{minipage}[htbp]{0.48\columnwidth}
    \centering
    \makeatletter\def\@captype{table}\makeatother\caption{Ablation Study}
    \vspace{-3mm}
    \resizebox{!}{0.55in} {
        \begin{tabular}{l c}
        \hline
        Ablation Module       & AUC     \\ 
        \hline
        Auxiliary Loss        & 0.6086  \\
        Other Attributes      & 0.6073  \\
        User Similarity       & 0.6073  \\
        Item Similarity       & 0.6065  \\
        Shallow Union         & 0.6092  \\
        Deep Union            & 0.6079  \\
        \hline
        \end{tabular}
    }   
\end{minipage}
\end{minipage}

\subsection {Comparison Experiments}
The offline comparison results are presented in Table 2 and the major observations are summarized as follows. 
\textbf{1)} DHAN employed hierarchical structures to capture important attribute information and outperformed traditional sequential modeling methods such as DIN and DIEN, which indicated the efficacy of utilizing attribute information.
\textbf{2)} R3S+DIN achieved 4.54\% improvement over DIN by extracting interaction features between the target and the trigger. DIHN captured users' intent on the trigger and achieved runner-up performance. Both of them highlight the importance of trigger information.
\textbf{3)} DPAN uses BCR to learn precise representations of similarity and diversity over attribute-aware values generated by AAVG and then uses SDUF to dynamically capture users' preferences under varying conditions, achieving the best result. 
Compared with DIHN focusing on the extent of users’ interests in the trigger to generate users’ dynamic interests representations, DPAN emphasizes users’ preferences for recommendations’ diversity and similarity under various conditions and then fuses these representations dynamically.

In our online relevant recommendation scenario, we conducted a rigorous half-month A/B testing with roughly equal traffic for both experiments, reaching over 600,000 users daily. As a result, DPAN outperformed DIN with a significant 7.62\% increase in CTR.

\subsection {Ablation Study}
Results in Table 3 demonstrate the impact of removing each module. 
\textbf{1) Auxiliary Loss of AAVG}: Removing the auxiliary losses resulted in a decline in AUC indicating that these losses help in better learning of embeddings for each attribute. \textbf{2) Attributes of AAVG}: To demonstrate the importance of multiple attributes, we reserved only the item as a basic attribute. The reduction in AUC highlights the necessity to leverage rich attribute information. \textbf{3) Similarity Representations of BCR}: With diversity representations reserved as the fundamental information for sequential recommendations, we investigate the influences of similarity representations of user interests and item information by removing each separately. The noticeable decline of the AUC underscores the importance of these similarity representations in capturing the user's instant interest, which is necessary for satisfying their preferences. \textbf{4) Union Networks of SDUF}: We analyzed the effect of shallow and deep union networks by removing each separately. Both networks provide additional information for learning users' preferences, and the deep union network had a relatively larger impact.

\subsection {Case Study}

\begin{figure}[htbp]
\vspace{-0.4cm}
\setlength{\abovecaptionskip}{0.2cm}
\setlength{\belowcaptionskip}{-0.4cm}
\includegraphics[scale=0.46]{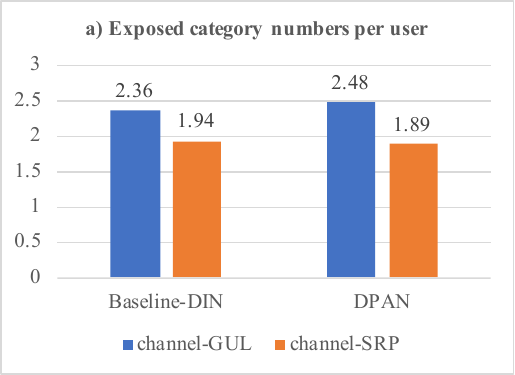}  
\hspace{0em}
\includegraphics[scale=0.46]{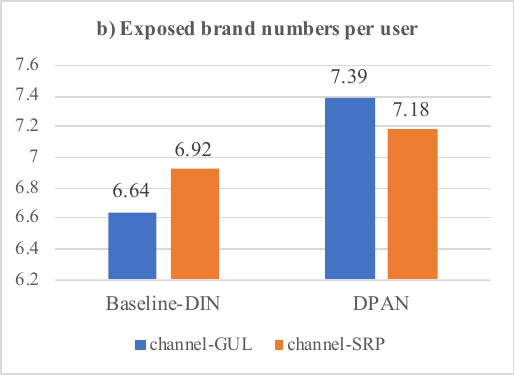}
\caption{Exposed category and brand numbers per user when users click from different channels.}
\label{fig:case}    
\end{figure}

In this section, we measure the model's ability to capture users' preferences for recommendation results' similarity and diversity by computing the exposed category and brand numbers per user. As there are many conditional factors influencing users' dynamic preferences, our analysis primarily focuses on the variations observed under different channels. As shown in Figure~\ref{fig:case}, DPAN is capable of providing items with a wider range of categories and brands when users navigate from GUL. This aligns with our expectations, as users clicking from GUL are more likely to seek out diverse options, whereas those clicking from SRP typically have a specific purchase in mind and tend to compare similar items, and DPAN's recommendations become more convergent in this case. In contrast, DIN, the baseline model, fails to provide such dynamic recommendations, resulting in inferior performance.

\section{conclusion}
In this paper, we present the Dynamic Preference-based and Attribute-aware Network (DPAN) for relevant recommendations. Through attribute-aware Bi-dimensional Compression-based Re-expression and Shallow and Deep Union-based Fusion, DPAN demonstrates its superiority in satisfying users’ dynamic preferences for recommendation results' similarity and diversity under various conditions. Extensive offline experiments and online A/B testing confirm the effectiveness of DPAN for relevant recommendations.








\end{sloppypar}
\end{document}